# Benchmarking acid and base dopants with respect to enabling the ice V to XIII and ice VI to XV hydrogen-ordering phase transitions

Alexander Rosu-Finsen and Christoph G. Salzmann[*]

Department of Chemistry, University College London, 20 Gordon Street, London WC1H 0AJ, United Kingdom; E-mail: c.salzmann@ucl.ac.uk

**Abstract**

Doping the hydrogen-disordered phases of ice V, VI and XII with hydrochloric acid (HCl) has led to the discovery of their hydrogen-ordered counterparts ices XIII, XV and XIV. Yet, the mechanistic details of the hydrogen-ordering phase transitions are still not fully understood. This includes in particular the role of the acid dopant and the defect dynamics that it creates within the ices. Here we investigate the effects of several acid and base dopants on the hydrogen ordering of ices V and VI with calorimetry and X-ray diffraction. HCl is found to be most effective for both phases which is attributed to a favourable combination of high solubility and strong acid properties which create mobile $H_3O^+$ defects that enable the hydrogen-ordering processes. Hydrofluoric acid (HF) is the second most effective dopant highlighting that the acid strengths of HCl and HF are much more similar in ice than they are in liquid water. Surprisingly, hydrobromic acid doping facilitates hydrogen ordering in ice VI whereas only a very small effect is observed for ice V. Conversely, lithium hydroxide (LiOH) doping achieves a performance comparable to HF-doping in ice V but it is ineffective in the case of ice VI. Sodium hydroxide, potassium hydroxide (as previously shown) and perchloric acid doping are ineffective for both phases. These findings highlight the need for future computational studies but also raise the question why LiOH-doping achieves hydrogen-ordering of ice V whereas potassium hydroxide doping is most effective for the 'ordinary' ice I$h$.

**Introduction**

The 'ordinary' hexagonal ice, ice I$h$, is well-known to display orientational disorder of its fully hydrogen-bonded water molecules.[1, 2] It is therefore strictly speaking not a truly crystalline material since translational symmetry is only present for the oxygen atoms. The disorder with respect to the hydrogen atoms gives rise to a molar configurational entropy that was estimated as $R \ln{^3/_2}$ by Pauling[1] and later confirmed experimentally.[3] All of the high-pressure phases of ice that can be crystallised from liquid water also display hydrogen disorder. This includes ices III, V, VI and VII, which have regions of thermodynamic stability in the phase diagram, as well as the two metastable ices IV and XII.[4] Interestingly, ices III and V are not fully hydrogen-disordered which means that some orientations of the water molecules are found more frequently than others.[5, 6]

Upon cooling, the hydrogen-disordered phases of ice are expected to undergo exothermic phase transitions to their hydrogen-ordered counterparts. This means that the overall connectivity of the hydrogen-bonded network remains unchanged but that orientational order of the water molecules is established at low temperatures as required by the 3$^{rd}$ law of thermodynamics. A fully hydrogen-ordered phase of ice displays zero configuration entropy which means that the Pauling entropy also reflects the expected entropy change for the transition from a fully hydrogen-ordered phase of ice to its fully hydrogen-disordered counterpart. Although, it should be noted that the Pauling entropy is strictly speaking only valid for a hypothetical dendritic ice structure without any rings. In reality, the configurational entropies of the various hydrogen-disordered ices are typically a few percent smaller.[7]

For ices III and VII, the hydrogen-ordering phase transitions to their hydrogen-ordered counterparts IX and VIII are readily observed upon cooling.[5, 6, 8-15] The requirement for a hydrogen-ordering phase transition to take place is that molecular reorientations are sufficiently fast in a hydrogen-disordered phase at the ordering temperature. Due to the fully hydrogen-bonded nature of water molecules in ice, molecular reorientation is in general a complicated and cooperative process which is only possible with the aid of mobile point defects. These include ionic $H_3O^+$ and $OH^-$ defects as well as D and L Bjerrum defects where either two or zero hydrogen atoms are found along a hydrogen bond.[16] In case of ices III and VII, the intrinsic point defects appear to be sufficient so that hydrogen ordering can be readily observed upon cooling.[8, 11] However, for all other hydrogen-disordered phases of ice including ices I$h$, I$sd$, IV, V, VI and XII, the molecular reorientation kinetics 'freeze-in' before reaching the ordering temperatures. So, instead of undergoing an exothermic phase transition to their hydrogen-ordered counterparts, these ices form orientational glasses at a glass transition temperature that lies above the ordering temperature.[4, 17-21] The $\Delta C_p$ values of these glass transitions are typically between $1 - 2$ J K$^{-1}$ mol$^{-1}$.[4, 17, 18, 21]

Doping ice with inorganic acids and bases provides a way to introduce extrinsic point defects in ice and this can speed-up molecular reorientation processes. It has been shown that potassium hydroxide (KOH) doped ice I$h$ undergoes a phase transition to ferroelectric ice XI below 72 K.[22-32] However, it is noted that the ferroelectric nature of ice XI has also recently been questioned.[33] Furthermore, complete hydrogen order has so far not been achieved for ice XI and the maximal degree of order has been estimated at around 60% on the basis of neutron diffraction data.[25, 30] Sodium hydroxide (NaOH) and rubidium hydroxide (RbOH) were also shown to promote hydrogen ordering of ice I$h$ although not to the same extent as KOH[23, 24] whereas hydrofluoric acid (HF) and barium hydroxide (Ba(OH)$_2$) were found to be ineffective.[24, 34, 35]

Salzmann *et al.* found that doping with hydrochloric acid (HCl) is highly effective in enabling ices V, XII and VI to become hydrogen ordered at low temperatures and the newly discovered hydrogen-ordered counterparts were named ices XIII, XIV and XV, respectively.[36-38] For ice IV, the HCl-doping was found to be significantly less effective. Analysis of calorimetric data showed that only about 5% of the Pauling entropy are lost around 115 K upon slow-cooling at 0.8 GPa.[4] This illustrates that ice samples can not only drop out of equilibrium before the ordering temperature is reached but also during the ordering transition. Therefore, the final degree of hydrogen order depends on the nature and concentration of the dopant, the cooling rate as well as on the pressure at which a doped ice sample is cooled.[4] A schematic summary of the various processes that can take place upon isobaric cooling of pure and doped ices is shown in Figure 5 of ref. [4].

The hydrogen-bonded network of ices V/XIII is shown in Figure 1(a). With 28 water molecules in the monoclinic unit cell this network structure is the most complicated amongst the currently known ices containing 4, 5, 6, 8, 9, 10 and 12-membered hydrogen-bonded rings.[4] Upon hydrogen ordering, a change of space group symmetry is observed from *A*2/*a* to *P*2$_1$/*a* and slow-cooling at ambient pressure produces an essentially fully hydrogen-ordered ice XIII.[36, 37, 39] A calorimetric study has shown that HCl doping is more effective than HF doping in enabling the hydrogen-ordering phase transition whereas KOH doping was found to be ineffective.[19] At ambient pressure, the reversible ice V ↔ ice XIII phase transition takes place at around 112 K over a 15 K temperature range and in at least two distinct but overlapping processes.[19, 36, 37] The loss in Pauling entropy upon hydrogen ordering has been estimated at 66%[19] which highlights the fact that ice V is already partially hydrogen ordered before undergoing the ice V → ice XIII phase transition.[6] The effectiveness of HCl doping in accelerating molecular reorientations in ice V was recently confirmed by dielectric-spectroscopy measurements in terms of the dielectric relaxation times.[40] Furthermore, the ice V ↔ ice XIII phase transition using HCl doping has also been followed with Raman[41] and 2D IR spectroscopy.[42]

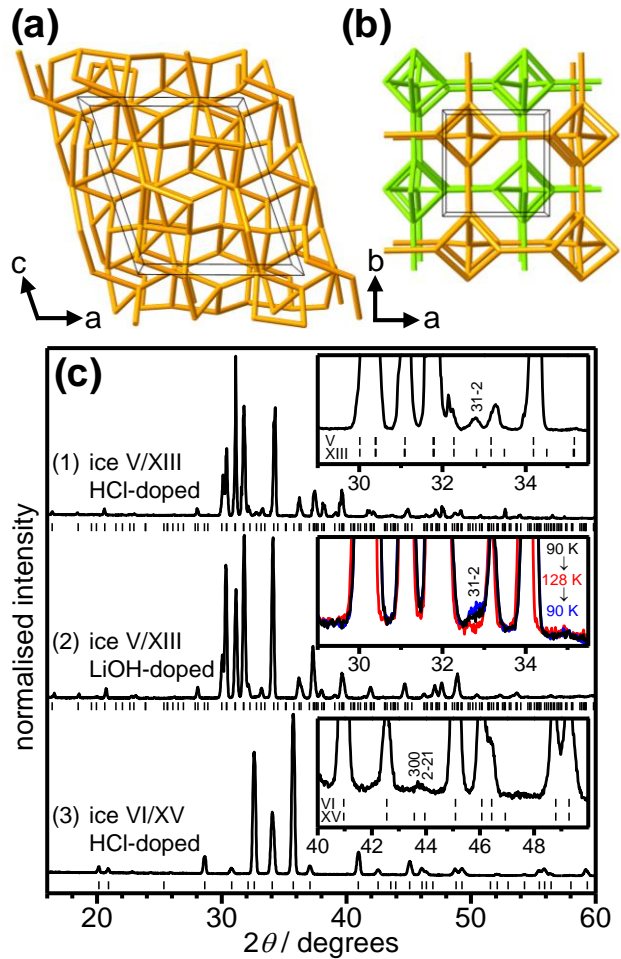

*Fig. 1* *Hydrogen-bonded networks of (a) ice V/XIII and (b) ice VI/XV. The nodes represent the oxygen atoms and lines the hydrogen bonds. The unit cell of ice V/XIII is shown with the centring of the P2$_1$/a space group of ice XIII. The two different hydrogen-bonded networks in ice VI/XV are highlighted in green and orange, respectively. The structures in (a) and (b) are shown on the same scale. (c) Powder X-ray diffraction at 95 K and ambient pressure of (1) HCl-doped ice V/XIII, (2) LiOH-doped ice V/XIII and (3) HCl-doped ice VI/XV after quenching under pressure. Tickmarks indicate the expected positions of Bragg peaks for ice XIII and ice XV in the main figure. The insets highlight the appearances of weak Bragg peaks which are not found for the hydrogen-disordered phases as indicated by the tick marks. In case of LiOH-doped ice V, the inset shows diffraction patterns at 95 K after quenching under pressure, after heating to 128 K at ambient pressure and after cooling back to 95 K at 1 K min$^{-1}$.*

Ices VI and XV contain two hydrogen-bonded networks that are not hydrogen-bonded to one another as shown in Figure 1(b).[38, 43, 44] Upon cooling at ambient pressure, HCl-doped ice VI first undergoes a sharp exothermic transition centred at about 130 K which is followed by a broad 'shoulder' reaching down to 100 K.[20] This ordering behaviour is also seen from the changes in lattice constants.[44] It has been suggested that the initial process could be due to independent ordering of the two networks followed by 'inter-network' hydrogen ordering at lower temperatures.[20] This scenario could potentially explain the slight ferroelectric behaviour observed for pure ice VI at low temperatures.[45] Slow-cooling DCl-doped ice VI leads to a change in space group symmetry from *P*4$_2$/*nmc* to *P*-1 and hence to an anti-ferroelectric structure.[36, 37, 43, 44] The loss in Pauling entropy during this transition is around 50% indicating that ice XV still contains considerable hydrogen-disorder[20] which is consistent with the fractional occupancies of the hydrogen sites determined from neutron diffraction.[38, 44] The ice VI ↔ ice XV phase transition was also followed using Raman spectroscopy.[46]

HCl-doped ice XII undergoes a phase transition upon cooling to ice XIV with a change in space group symmetry from $I$-$42d$ to $P2_12_12_1$.[36, 37, 47] The ice XII ↔ ice XIV phase transition was found to show some reversibility upon slow-cooling at ambient pressure.[36, 37, 48] However, a much less ordered ice XIV was obtained. So, contrary to the behaviours of ices XIII and XV, ice XIV is easier to order at higher pressures. It was proposed that the pressure helps overcome orthorhombic strain which arises during the ice XII → ice XIV phase transition.[36, 37] On the basis of calorimetric data, the main claim of ref. [49] was a full loss of Pauling entropy upon cooling HCl-doped $H_2O$ ice XII at cooling rates smaller than 15 K min$^{-1}$ at 0.8 GPa implying that a fully hydrogen-ordered ice XIV had been obtained. However, our integration of the calorimetry data provided in Figure 1(a) of ref. [49] only gives a change of 51 % of the Pauling entropy which means that even slow-cooling under pressure leaves some residual hydrogen disorder in ice XIV. Consistent with this, the Raman spectra of $H_2O$ ice XIV obtained after cooling at 1 K min$^{-1}$ at 1.2 GPa do not show the spectroscopic characteristics expected for a fully hydrogen-ordered phase of ice.[48]

The exact mechanisms of the hydrogen-ordering processes in the various ices are still unclear.[16] Closely tied to this is the question how acid and base dopants accelerate molecular reorientation processes, and what the factors are that govern the effectiveness of a specific dopant.[4] In the case of HCl doping, it has been suggested that the doping creates mobile $H_3O^+$ point defects that leave trails of water molecules with changed orientations behind as they migrate through the hydrogen-bonded networks and thus facilitate molecular reorientations within the entire crystal. The effectiveness of a dopant therefore depends not only on its concentration within an ice crystal but also on the mobility of the created point defects.[41] In ref. [19] we attributed the higher effectiveness of HCl *versus* HF doping in achieving hydrogen-ordering in ice XIII to the greater acid strength of HCl and therefore higher tendency to form travelling $H_3O^+$ defects.

Here we test and compare a large range of acid and base dopants with respect to their abilities to facilitate hydrogen-ordering of ices V and VI at ambient pressure including hydrofluoric, hydrochloric, hydrobromic and perchloric acid as well as lithium, sodium and potassium hydroxide. The aim is to gain more detailed insights into the dopant solubilities and defect dynamics in those ices but also to investigate if more effective dopants than hydrochloric acid can be found.

**Experimental Methods**

*Preparation of doped ice V/XIII and VI/XV samples*

HCl and perchloric acid ($HClO_4$) as well as NaOH solutions were purchased as 0.01 M standard solutions from Sigma Aldrich. Concentrated HF and hydrobromic acid (HBr) were diluted with MilliQ water to obtain 0.01 M solutions and the target concentration was confirmed by titration with 0.01 M

NaOH standard solution. A 0.01 M lithium hydroxide (LiOH) solution was prepared by dissolving LiOH powder in MilliQ water and the target concentration was confirmed by titration with 0.01 M HCl standard solution.

For the preparation of doped ice V samples, 800 μL of the various solutions were quickly pipetted into an indium cup inside a 1.3 cm diameter hardened-steel pressure die precooled with liquid nitrogen. The sample was then pressurized with a 30 t hydraulic press to 0.5 GPa, heated isobarically to 250 K and quenched with liquid nitrogen while keeping the pressure constant. After reaching 77 K, the pressure was slowly released to ambient pressure. Finally, the sample was recovered under liquid nitrogen and freed from the indium.

Doped ice VI samples were prepared using 700 μL of solution, a 1.0 cm diameter pressure die and the samples were heated isobarically at 1.0 GPa to 250 K followed by quenching and the same recovery procedure as for the doped ice V samples.

*Powder X-ray diffraction*

Powders of the recovered ice samples were transferred under liquid nitrogen into a purpose-built Kapton-window sample holder mounted on a Stoe Stadi P diffractometer with Cu K$_{\alpha 1}$ radiation at 40 kV, 30 mA and monochromated by a Ge 111 crystal. Data were collected using a Mythen 1K linear detector and the temperature of the samples was maintained at 95 K with an Oxford Instruments CryojetHT.

*Differential Scanning Calorimetry*

A few small pieces of the doped high-pressure phases of ice were transferred into stainless-steel capsules with screwable lids under liquid nitrogen. These were quickly transferred into a precooled Perkin Elmer DSC 8000 advanced double-furnace differential scanning calorimeter with a base temperature of 93 K. Thermograms were recorded upon heating to 138 K at 5 K min$^{-1}$ followed by cooling back to 93 K at 0.5, 5 or 30 K min$^{-1}$. The samples were then heated at 5 K min$^{-1}$ from 93 to 263 K and the previous heating / cooling procedure was repeated with now ice I$h$. Finally, the moles of ice in the DSC capsules were determined by recording the endothermic melting of ice at 0°C, and using 6012 J mol$^{-1}$ as the enthalpies of melting of H$_2$O ice I$h$. The thermograms of ice I$h$ were subtracted from the previously recorded data as a background correction. The resulting DSC signal was divided by the number of moles of H$_2$O and the heating / cooling rate which yields a quantity with J mol$^{-1}$ K$^{-1}$ as the unit.

**Results and discussion**

*Effects of acid and base dopants on the ice V to XIII hydrogen-ordering phase transition*

Consistent with our earlier study,[19] pressure-quenched HCl-doped ice XIII shows a strong endotherm around 115 K upon heating as shown by thermogram (1a) in Figure 2. This phase transition corresponds to the hydrogen-disordering phase transition to ice V which is reversible upon cooling back as can be seen from the exotherm in thermogram (1b). The sample identity was confirmed by X-ray diffraction as shown in Figure 1(c). Remarkably, the (31-2) Bragg peak, which is characteristic for ice XIII,[36] can be observed in the diffraction pattern. Since X-rays are mainly scattered by the oxygen atoms in ice, the appearance of this peak indicates the slight displacement of the oxygen atoms away from the positions defined by the *A2/a* space group symmetry of ice V (*i.e.* Wyckoff splitting).

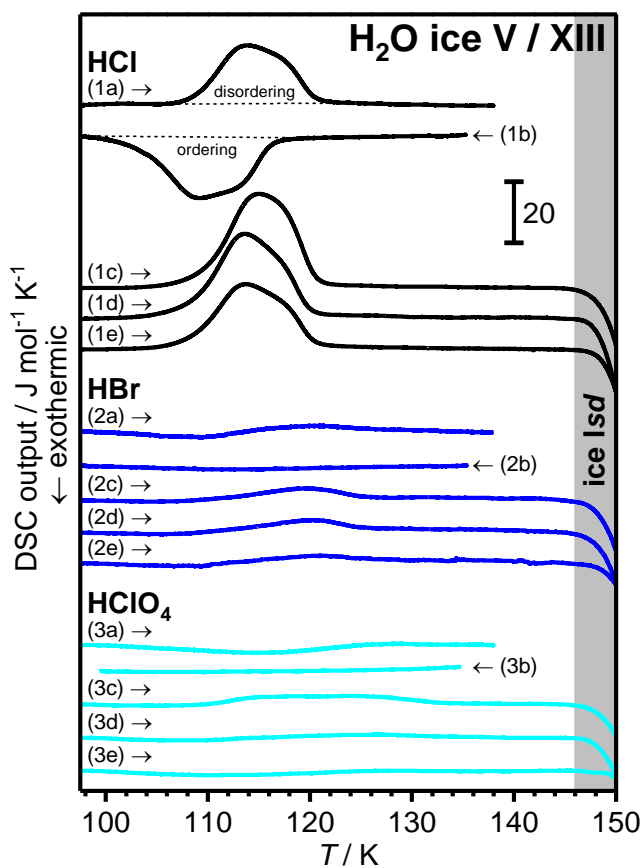

*Fig. 2* DSC thermograms of HCl (black), HBr (blue) and HClO$_4$-doped (cyan) ice V samples recorded upon heating from 93 K to 138 K at 5 K min$^{-1}$ (1a, 2a, 3a) followed by cooling at 5 K min$^{-1}$ back to 93 K (1b, 2b, 3b). Thermograms (1c, 2c, 3c), (1d, 2d, 3d) and (1e, 2e, 3e) were recorded upon heating at 5 K min$^{-1}$ from 93 K after cooling at 0.5, 5 and 30 K min$^{-1}$ from 138 K, respectively. The grey-shaded region indicates the temperature range of the irreversible crystallisation of ice V to ice Isd. The scale bar in this figure has the same size as in all the following DSC figures and the intensities are therefore directly comparable.

Thermograms (1c-e) were recorded after previous cooling from 138 K at 0.5, 5 and 30 K min$^{-1}$, respectively, and therefore contain information about the kinetics of the ice V → ice XIII phase transition. A more ordered ice XIII is expected to display a stronger endothermic feature upon the phase transition to ice V. The recorded enthalpies are shown as a function of the preceding cooling rate in Figure 3(a). As described earlier,[19] the expected shape of the curves in this plot is a sigmoid or "S"-shaped curve with a first plateau at very small cooling rates corresponding to the fully ordered state and a second one at zero enthalpy for the quenched disordered state. However, depending on the dopant in question, only a segment of this curve is accessible within the limits of the experimentally possible cooling rates. The newly determined enthalpies for HCl-doped ice V/XIII from this study agree well with the values from ref. [19].

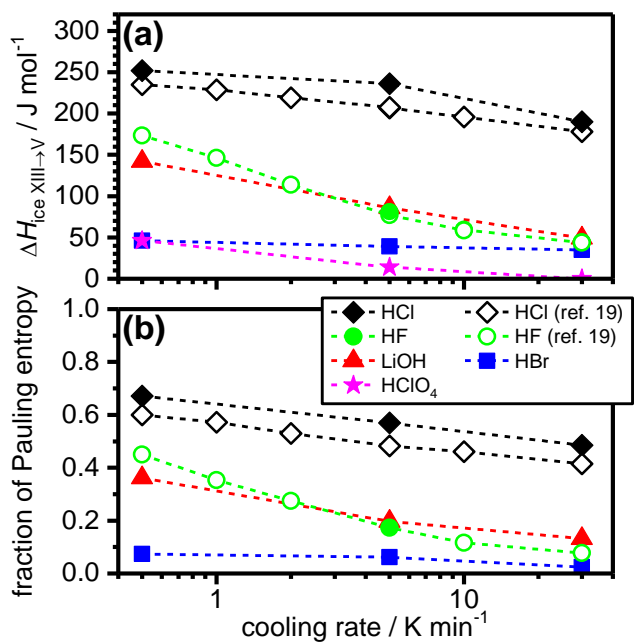

***Fig. 3*** *(a) Enthalpies of the ice XIII to ice V phase transition recorded upon heating after cooling doped ice V at the specified cooling rates. (b) Corresponding fractions of the Pauling entropy (3.371 J mol$^{-1}$K$^{-1}$).*

Since the ice XIII → ice V phase transition is reversible in the case of HCl-doped ice, it is possible to calculate the changes in entropy arising from the transition.[19, 20, 50] This is achieved by dividing the thermograms by the temperature and subsequent peak integration. In Figure 3(b), the determined entropy changes are shown as fractions of the Pauling entropy. As mentioned earlier, HF doping is also capable of facilitating the ice V → ice XIII phase transition albeit to a less effective degree. This can be seen from the corresponding enthalpy and entropy changes from ref. [19] which are shown together with one new data point in Figure 3.

Since the effectiveness of the dopants increases upon going from HF to HCl, the next hydrohalic acid in this sequence, HBr, was investigated. As shown in Figure 2, only very weak endothermic transitions are observed upon heating the HBr-doped samples and no exotherm is found upon cooling. The corresponding enthalpies are shown in Figure 3(a). As mentioned earlier, entropy values can only be obtained if the calorimetric features in question are reversible. Also, great care needs to be taken that endothermic relaxation features related to the glass transitions of the hydrogen-disordered ices are not used for the calculation of entropy changes. Such endothermic features do not produce latent heat but are kinetic in origin.[17, 19, 20] Arguing that the endothermic features of the HBr-doped sample seem to appear in a reversible fashion and in the temperature range of the ice XIII ↔ ice V phase transition, the corresponding entropy values are shown in Figure 3(b). However, this of course does not change the fact that the overall effect of HBr doping is very small.

Perchloric acid is one of the strongest inorganic acids (*cf.* Table 1) and we therefore also investigated its effect on the ice V → ice XIII phase transition. As shown in Figure 2, again only very weak endothermic features were observed. The most pronounced endotherm was recorded after cooling at 0.5 K min$^{-1}$ at ambient pressure. However, since only about half of the endotherm is located within the temperature range of the ice XIII ↔ ice V phase transition, entropy values were not calculated for this transition.

As mentioned earlier, the effectiveness of an acid dopant in facilitating a hydrogen-ordering process is expected to depend on its solubility in ice as well as on the ability of the acid to dissociate within the ice and create a mobile $H_3O^+$ defect. The solubility of HCl in ice I$h$ has been determined with a mole fraction of $3\times10^{-5}$.[51] The solubility of HF is slightly larger with a mole fraction between $4\times10^{-5}$ and $5\times10^{-5}$.[52, 53] For our doping experiments, we typically use 0.01 M acid solutions as the initial starting solutions (*cf.* Experimental Section) which corresponds to a mole fraction of the acids of $1.8\times10^{-4}$. At present, there are no solubility data of inorganic acids in any of the high-pressure phases of ice in the literature. However, the way the enthalpy values of, for example, the ice XIII → ice V transition after cooling at 0.5 K min$^{-1}$ change as a function of the HCl concentration in the initial sample solution should give some measure for the solubility of HCl in ice V/XIII. The argument here is that an excess of HCl should not lead to more efficient hydrogen ordering within a crystal. Upon decreasing the mole fraction of HCl from $1.8\times10^{-4}$ to $1.8\times10^{-5}$ a 14% decrease of the transition enthalpy was observed whereas lowering the mole fraction to $1.8\times10^{-6}$ gave an 80% decrease in the enthalpy.[19] It is therefore reasonable to conclude that the maximal mole fraction of HCl in ice V/XIII is somewhere between $1.8\times10^{-4}$ and $1.8\times10^{-5}$, and probably closer to $1.8\times10^{-5}$. This implies that the solubility of HCl in ice V/XIII is similar to its solubility in ice I$h$. In fact, it is even possible that the concentration of HCl within ice V/XIII is limited in our experiments by the solubility of HCl in ice I$h$ which forms first as the initial sample solutions are frozen.

The higher density of ice V/XIII does therefore not necessarily mean that acids and perhaps other species as well are less soluble than in ice I. In fact, it is noted that the intermolecular hydrogen-bonded distances increase in length in the high-pressure phases compared to ice I,[4] which could favour the incorporation of ionic species. In fact, it has been shown that alkali halides can dissolve in the very dense ice VII.[54, 55] For HCl in ice I$h$, it has been suggested that the chloride anions substitute water molecules within the ice and therefore experience the hydrogen-bonding environment from the neighbouring water molecules.[56] However, the occupation of interstitial sites may also be possible at higher temperatures as part of diffusion pathways.[51]

HCl, HBr and $HClO_4$ are generally considered to be strong acids that dissociate fully in liquid water as indicated by their p$K_A$ values (*cf.* Table 1). In contrast, HF is a weak acid that only shows a small degree of dissociation in water.[57] With respect to the dissociation of inorganic acids in ice, experimental studies suggests that HCl, HBr and $HClO_4$ dissociate readily at ice surfaces.[58-60] According to DFT calculations, the dissociation of HCl located within ice I is a barrierless process whereas the dissociation of HF is substantially hindered.[61] On the basis of elegant fluorescence-quenching experiments, a wide range of weak and strong acids were classified with respect to their acid strength in ice I$h$.[62, 63] The general trend is that strong acids release more protons ($H^+$) into ice than weak acids. Yet, the differences between strong and weak acids seem to be much less pronounced compared to the situation in liquid water. Strong acids only release about 3 times more protons into ice than typical weak acids. At around 1 mM concentrations, the degree of dissociation of weak acids is about 30% whereas strong acids are completely dissociated. HF seems to go slightly against the general trend and has been found to be a slightly stronger acid in ice compared to other weak acids that are more acidic in water. Assuming that a similar behaviour is valid for ice V/XIII, these results are consistent with our measurements in the sense that HCl is more effective than HF in facilitating the ice V ↔ ice XIII phase transitions but certainly not by many orders of magnitude as the $pK_A$ values in liquid water would suggest. This also means that the dramatic drop in efficiency observed for HBr and $HClO_4$ must be due to the low solubilities of these species in ice. The ionic radius of $Cl^-$ is 40% greater compared to $F^-$, whereas $Br^-$ and $ClO_4^-$ are 53% and 160% larger, respectively (*cf.* Table 1). This suggest that there must be a cut-off point between the ionic radii of $Cl^-$ and $Br^-$ above which the incorporation into ice V/XIII becomes very unfavourable. The exact details of the electronic interactions of the anions with the ice matrix will of course also be important. Amongst the acids investigated, the high efficiency of HCl in facilitating the ice V ↔ ice XIII phase transition can be attributed to a combination of high solubility and strong acid properties.

Not much is known about the solubilities of bases including alkali hydroxides in ice. In liquid water, LiOH, NaOH and KOH are all considered to be strong bases and fully dissociated as reflected by their $pK_B$ values (*cf.* Table 1). So, it seems reasonable to assume that their basicities remain roughly similar upon incorporation into ice. However, surprisingly, LiOH doping was found to facilitate the ice V ↔ ice XIII phase transition as indicated in Figure 4 by the endotherms upon heating and the exotherm upon cooling. As observed for HCl and HF-doping, slow-cooling at ambient pressure leads to a more ordered ice XIII. Remarkably, this process can also been seen in X-ray diffraction as shown in the inset in Figure 1(c). The (31-2) Bragg peak, which is characteristic for ice XIII, is present after pressure-quenching the sample, it disappears after heating 128 K consistent with the formation of ice V and reappears with a slightly higher intensity after cooling at 1 K min$^{-1}$ which is in agreement with the trends shown in Figure 4. Judging from the enthalpy and entropy values shown in Figure 3, the performance of LiOH-doping is comparable with HF-doping.

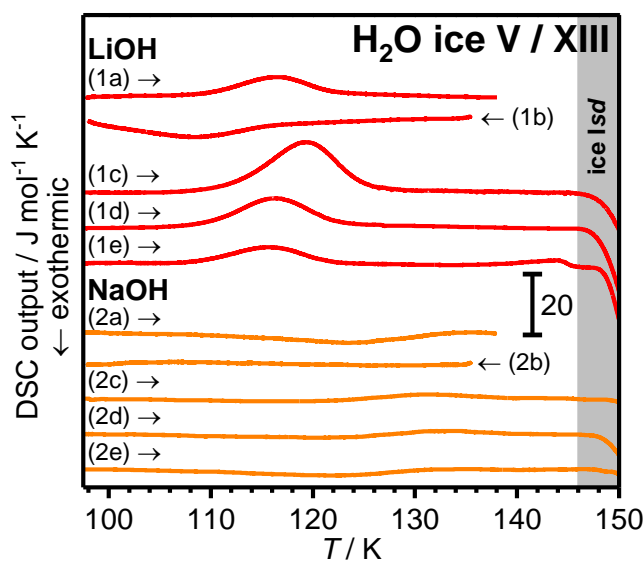

***Fig. 4*** *DSC thermograms of LiOH (red) and NaOH-doped (orange) ice V samples recorded upon heating from 93 K to 138 K at 5 K min$^{-1}$ (1a, 2a) followed by cooling at 5 K min$^{-1}$ back to 93 K (1b, 2b). Thermograms (1c, 2c), (1d, 2d) and (1e, 2e) were recorded upon heating at 5 K min$^{-1}$ from 93 K after cooling at 0.5, 5 and 30 K min$^{-1}$ from 138 K, respectively.*

The NaOH-doped sample on the other hand shows only weak endothermic features upon heating which are consistent with the glass transition of ice V and no exothermic transition is observed upon cooling. Similar behaviour has been previously observed for KOH-doped ice V.[19] Although, it should be noted that the KOH-doped sample in ref. [19] was slow-cooled at 0.5 K min$^{-1}$ under pressure and therefore it exhibited a more pronounced endothermic relaxation feature than the one observed for the NaOH-doped sample in Figure 4. In any case, neutron diffraction has shown that a slow-cooled KOD-doped sample did not show signs of hydrogen order.[36]

On the basis of similar basicities of the three alkali hydroxides, it seems reasonable to attribute the ability of LiOH to facilitate the ice V ↔ ice XIII phase transition to a higher solubility. The ionic radius increases by 32% from Li$^+$ to Na$^+$, and by 72% from Li$^+$ to K$^+$ (*cf.* Table 1). This means that the incorporation of monovalent cations with ionic radii larger than Li$^+$ into ice V/XIII is a difficult process.

*Effects of acid and base dopants on the ice VI to XV hydrogen-ordering phase transition*
The same acid and base dopants as in the previous section are now tested with respect to their abilities facilitating the ice VI ↔ ice XV phase transition at ambient pressure. Consistent with the thermograms shown in ref. [20], pressure-quenched HCl-doped ice VI/XV shows an irreversible hydrogen-ordering process upon heating starting above 100 K, followed by hydrogen-disordering at higher temperatures to ice VI (*cf.* Figure 5). It is noteworthy that the DCl-doped D$_2$O ice VI/XV sample in ref. [20] did not show such 'transient' ordering in DSC upon heating at 10 K min$^{-1}$. Yet, it was observed for a deuterated sample upon heating at 0.4 K min$^{-1}$ as indicated by the changes in lattice constants and the appearance of Bragg peaks characteristic for ice XV.[44] This means that the appearance of the 'transient'-ordering feature is strictly determined by kinetic factors. In fact, such 'transient' ordering has also been seen for ice V/XIII from the changes in lattice constants upon slowly heating partially disordered ice XIII.[37] Ice XV is obtained at ambient pressure after cooling from 138 K with a broad exothermic feature that reaches down to about 100 K.

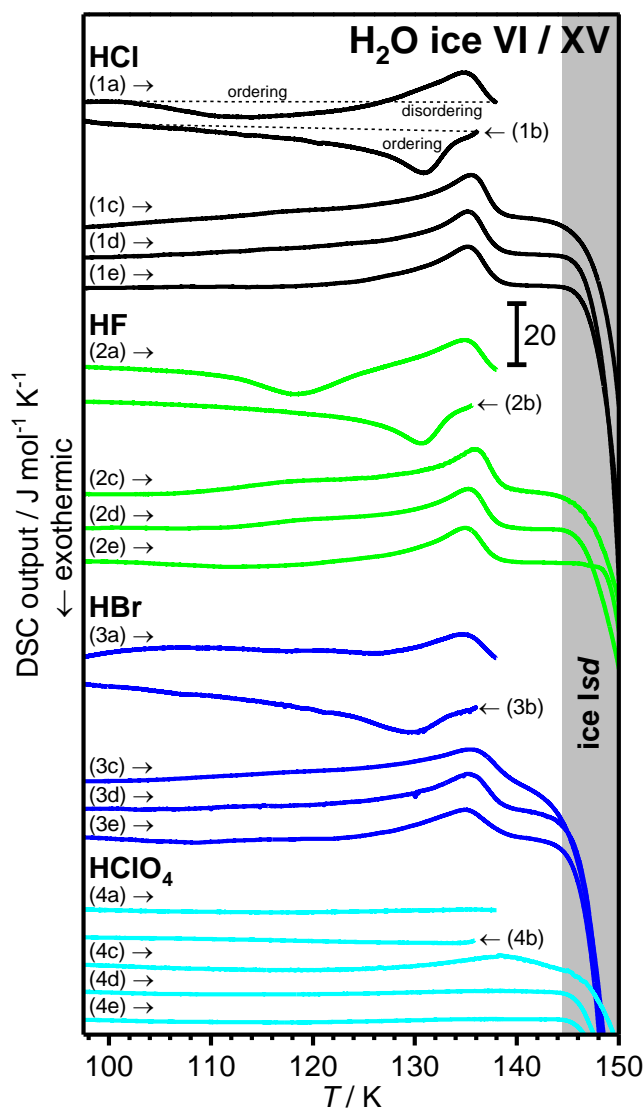

*Fig. 5* DSC thermograms of HCl (black), HF (green), HBr (blue) and HClO$_4$-doped (cyan) ice VI samples recorded upon heating from 93 K to 138 K at 5 K min$^{-1}$ (1a, 2a, 3a, 4a) followed by cooling at 5 K min$^{-1}$ back to 93 K (1b, 2b, 3b, 4b). Thermograms (1c, 2c, 3c, 4c), (1d, 2d, 3d, 4d) and (1e, 2e, 3e, 4e) were recorded upon heating at 5 K min$^{-1}$ from 93 K after cooling at 0.5, 5 and 30 K min$^{-1}$ from 138 K, respectively. The grey-shaded region indicates the temperature range of the irreversible crystallisation of ice VI to ice Isd.

Figure 1(c) shows the X-ray diffraction pattern of a pressure-quenched HCl-doped ice VI/XV sample which displays weak (300) and (2-21) Bragg peaks characteristic of ice XV.[38, 44] As previously mentioned for ice XIII, these do not arise from the ordering of the hydrogen sites but from the displacement of the oxygen atoms away from the positions in hydrogen-disordered ice VI.

The enthalpy values of the ice XV → ice VI phase transition recorded upon heating HCl-doped ice XV samples as a function of the previous cooling rate are in good agreement with the values from ref.

[20] (*cf.* Figure 6). The maximal loss in Pauling entropy is found just below 50% after slow-cooling at 0.5 K min$^{-1}$.

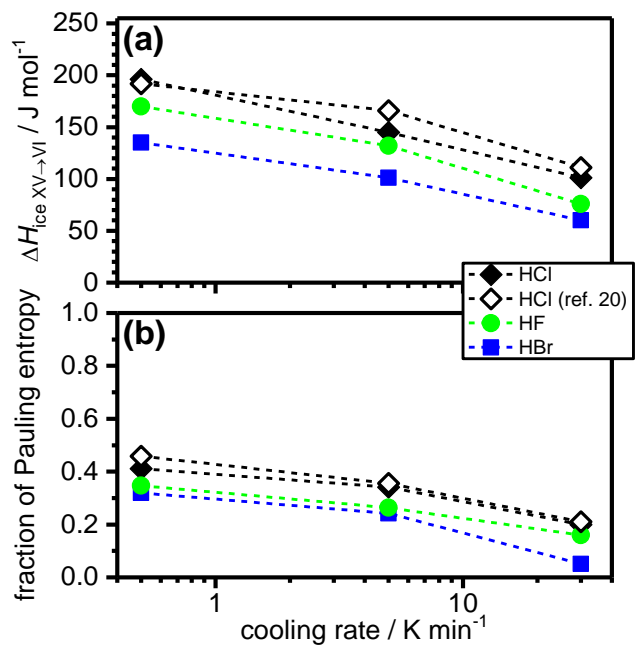

**Fig. 6** *(a) Enthalpies of the ice XV to ice VI phase transition recorded upon heating after cooling doped ice VI at the specified cooling rates. (b) Corresponding fractions of the Pauling entropy (3.371 J mol$^{-1}$K$^{-1}$).*

Although not as effective as HCl-doping, HF-doping also achieves hydrogen ordering of ice VI as indicated by the thermograms shown in Figure 5, and the enthalpy and entropy values in Figure 6. The gap between the HCl and HF data in Figure 6 is smaller in particular at higher cooling rates compared to the data shown in Figure 3 for the ice V ↔ ice XIII phase transition. This indicates that HF-doping is relative to HCl-doping more effective for the ice VI ↔ ice XV phase transition than for ice V ↔ ice XIII.

Remarkably, HBr-doping also achieves considerable hydrogen-ordering in ice VI which is in clear contrast to ice V (*cf.* Figures 2 and 3). The efficiencies of the three hydrohalic acids in enabling the ice VI ↔ ice XV phase transition are therefore HCl > HF > HBr. This sequence is also reflected in the shifts of the onset temperatures of the 'transient' ordering features recorded upon first heating of the samples. In agreement with the situation for ice V, HClO$_4$-doping does not appear to induce any hydrogen ordering as evidenced by the absence of an exothermic feature upon cooling. The best performance of HCl-doping can again be explained by a favourable combination of good solubility and strong acid properties. Considering that both HCl and HBr are strong acids, the poorer performance of HBr is attributed to a somewhat lower solubility of Br$^-$ in ice V/XV compared to Cl$^-$.

As shown in Figure 7 and Figure 2 of ref. [20], LiOH, NaOH and KOH do not facilitate the ice VI ↔ ice XV phase transition. This means that the alkali hydroxides are either insoluble in ice VI or alternatively that the mechanism of hydrogen ordering in ice VI is somehow incompatible with base doping. As for NaOH and KOH-doped ice V, the weak endothermic features recorded upon first heating are attributed to relaxation features associated with the glass transitions observed for the pure hydrogen-disordered ices.

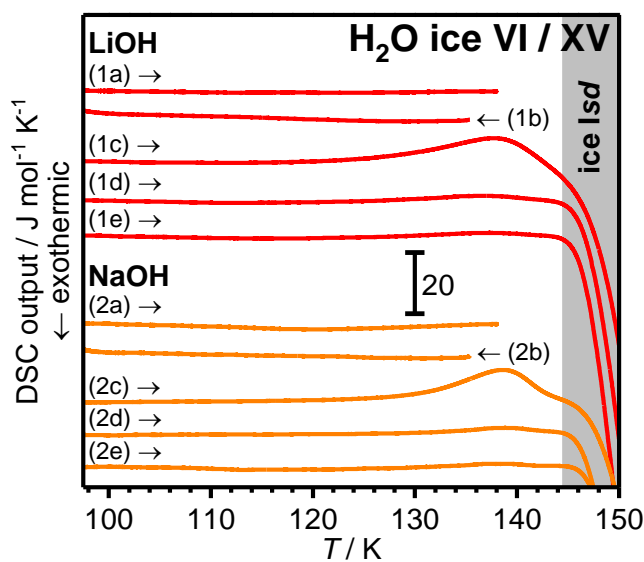

*Fig. 7* DSC thermograms of LiOH (red) and NaOH-doped (orange) ice VI samples recorded upon heating from 93 K to 138 K at 5 K min$^{-1}$ (1a, 2a) followed by cooling at 5 K min$^{-1}$ back to 93 K (1b, 2b). Thermograms (1c, 2c), (1d, 2d) and (1e, 2e) were recorded upon heating at 5 K min$^{-1}$ from 93 K after cooling at 0.5, 5 and 30 K min$^{-1}$ from 138 K, respectively.

**Conclusions**

Out of all the investigated acid and base dopants, HCl is most efficient in facilitating both the ice V ↔ ice XIII as well as the ice VI ↔ ice XV phase transition. This is attributed to a favourable combination of high solubility and strong acid properties which create the mobile $H_3O^+$ defects that enable the hydrogen-ordering processes. HF is the 'runner-up' in both cases. Base dopants as well as $HClO_4$ were generally found to show very poor performances for both phase transitions. However, the exception is LiOH-doping which displayed a similar efficiency in hydrogen-ordering ice V as HF-doping. The effectiveness of LiOH-doping compared to NaOH and KOH is explained by a smaller size of the cation and hence greater solubility in ice V. However, this raises an important question with respect to the ice I$h$ → ice XI phase transition for which KOH-doping was found to be most effective, and NaOH and RbOH-doping showed poorer performances in calorimetry.[23, 24] Consistent with this, the dielectric-relaxation times of alkali hydroxide-doped ice I$h$ at 100 K increase in the order KOH < NaOH < RbOH < LiOH.[64] The reasons for the different behaviours of ices I$h$ and V are unclear at present.

HBr-doping was found to achieve significant hydrogen-ordering of ice VI but not of ice V. Since ice VI/XV consists of two interpenetrating networks, it is intriguing to speculate that fragments of the individual networks can be replaced with Br$^-$ anions which is not possible for the more 'tightly-knit' network of ice V/XIII. In any case, this study highlights the urgent need for future computational work in this area aimed at determining the substitution energies of acid and base dopants into the various high-pressure phases of ice.

**Acknowledgements**

We thank the Royal Society for a University Research Fellowship (CGS, UF100144), M. Vickers for help with the X-ray diffraction measurements and J. Cockcroft for access to the Cryojet. This project has received funding from the European Research Council (ERC) under the European Union's Horizon 2020 research and innovation programme (grant agreement No 725271).

**Table 1.** p$K_A$ and p$K_B$ values of HF, HCl, HBr, HClO$_4$, LiOH, NaOH and KOH as well as the ionic radii of the corresponding anions and cations.[65-70]

| acid / base | p$K_A$ / p$K_B$ | ionic radii of anions and |
|---|---|---|
| HF | 3.17[65] | 1.19[69] |
| HCl | -6.3[66] | 1.67[69] |
| HBr | -9[67] | 1.82[69] |
| HClO$_4$ | -10[67] | 3.09[70] |
| LiOH | -0.04[68] | 0.88[69] |
| NaOH | -0.93[68] | 1.16[69] |
| KOH | -0.7[68] | 1.52[69] |